# Bayesian Inference with Spiking Neurons

Michael G. Paulin[1] and Andre van Schaik[2]

[1]Department of Zoology, University of Otago, New Zealand. [2]The MARCS Institute, University of Western Sydney.

**Abstract**

Humans and other animals behave as if we perform fast Bayesian inference underlying decisions and movement control given uncertain sense data. Here we show that a biophysically realistic model of the subthreshold membrane potential of a single neuron can exactly compute the numerator in Bayes rule for inferring the Poisson parameter of a sensory spike train. A simple network of spiking neurons can construct and represent the Bayesian posterior density of a parameter of an external cause that affects the Poisson parameter, accurately and in real time.

# 1   Introduction

Behavioural observations and psychophysical experiments show that humans and other animals act as if our decisions and actions are based on Bayesian inference (Kording and Wolpert, 2006; Kording, 2007; Laurens and Droulez, 2007; Stevenson et al., 2009; Berniker and Kording, 2011; Franklin and Wolpert, 2011; Wolpert and Landy, 2012; Pouget et al., 2013). There is compelling evidence that neurons in the central nervous system explicitly compute and represent Bayesian posterior probabilities of states and events, given observations provided by sensory neurons (Rao, 2006; Gold and Shadlen, 2007; MacNeilage et al., 2008; Angelaki et al., 2009; Lochmann and Deneve, 2011; Fischer and Pena, 2011; Karmali and Merfeld, 2012).

A number of models of neural mechanisms of Bayesian inference have been proposed (Wu and Amari, 2002; Paulin, 2005; Jazayeri and Movshon, 2006; Huys et al., 2007; Ma et al., 2008; Beck et al., 2008; Bobrowski et al., 2009; Buesing et al., 2011; Boerlin and Deneve, 2011). They mostly represent activity of spiking neurons using instantaneous firing rates or spike counts in time intervals. But Bayesian inference can be applied directly to point process observations without smoothing or binning (Brown et al., 1998; Wiener and Richmond, 2003; Deneve, 2008; Koyama et al., 2010; Deneve, 2012; Susemihl et al., 2013). This provides a mathematical framework for modeling how spikes themselves can be treated as observations in models of Bayesian inference in nervous systems (Paulin et al., 2004; Deneve, 2008, 2012; Paulin and Hoffman, 2011).

The Bayesian posterior given a stream of observations can be computed by representing it with a finite random sample that is updated to construct a new sample from the posterior when each new observation becomes available. This is called a particle filter (Doucet et al., 2001). Each sample point, representing a possible state of the world, is called a particle. In this paper we show that spiking neurons can infer world states from sensory spike trains accurately in real time using spikes as particles. The foundation of the model is a correspondence between Bayes rule for sequential inference from observations of a Poisson process, and the subthreshold membrane dynamics of a single neuron.

# 2   Bayes rule and the single neuron

If spikes are generated by a homogeneous Poisson point-process with intensity $\lambda$, then the probability that no spikes occur in an interval of duration $t$ is $e^{-\lambda t}$ and the probability that at least one spike occurs in this interval is $1 - e^{-\lambda t}$.

The probability that at least one spike occurs in a specified subinterval of duration $r < t$ and no other spikes occur in the interval is

$$L(\lambda, t, r) = e^{-\lambda(t-r)}\left(1 - e^{-\lambda r}\right) = e^{-\lambda t}\left(e^{\lambda r} - 1\right). \qquad (1)$$

Suppose that the spike train is censored so that spikes do not occur closer together than a small finite period of duration $r$. This may be either because the spike-generating mechanism cannot generate a second spike, or because the observing mechanism is not able to detect a second spike, within a refractory period of a preceding spike. Then $L(\lambda,t,r)$ is the likelihood for $\lambda$ at time $t$, if a spike occurred at time 0 and no more have occurred.

If the prior probability density $f(\lambda,0)$ is known at the spike time then the posterior density at time $t$ is, by Bayes rule,

$$f(\lambda, t) = \frac{e^{-\lambda t}\left(e^{\lambda r} - 1\right) f(\lambda, 0)}{\int_0^\infty e^{-\lambda t}\left(e^{\lambda r} - 1\right) f(\lambda, 0)\,\mathrm{d}\lambda} \tag{2}$$

This equation can be applied sequentially to estimate the posterior density of $\lambda$ from a spike train. Each time a spike occurs, $t$ is reset to 0 and the posterior at that time becomes the prior for inference in the interval before the next spike occurs.

At a point $\lambda^*$ the numerator in equation 2 is the passive response of a first-order linear low-pass filter,

$$\frac{dV(\lambda^*, t)}{dt} + \lambda^* V(\lambda^*, t) = 0 \tag{3}$$

with time constant $\tau^* = 1/\lambda^*$ and initial condition

$$V(\lambda^*, 0) = \left(e^{\lambda^* r} - 1\right) f(\lambda^*, 0) \tag{4}$$

If a spike arrives at time $t_1$ and the filter output is reset to

$$V(\lambda^*, t_1) = \left(e^{\lambda^* r} - 1\right) f(\lambda^*, t_1) \tag{5}$$

at that instant, then the filter's response after the spike is the Bayes numerator given the prior probability density *and* the spike. This can be repeated at subsequent spike times, so that when driven by a sequence of events from a Poisson process, the filter response is the Bayes numerator at $\lambda^*$, given prior density $f(\lambda^*, 0)$ and all spikes up to the current time.

The amplitude of the impulse required to reset the filter at the $j$th spike time is the difference between the desired initial condition (equation 5) and the pre-spike response of the filter (the numerator in equation 2),

$$\left(e^{\lambda^* r} - 2\right) f(\lambda^*, t_j) \tag{6}$$

If the probability of a spike in a refractory period is small ($\lambda r << 1$), then equation 6 indicates that the spike input must be inhibitory. The posterior probability is then obtained by uniformly re-scaling the filter's response, according to equation 2. Thus the shape of the posterior probability density is sculpted by spikes, while its scale is determined by normalization.

This analysis leads to a coupled pair of differential-algebraic equations for continuously inferring the posterior density of the Poisson parameter $\lambda$ of a spike train at a point $\lambda^*$,

$$\frac{dV(\lambda^*, t)}{dt} + \lambda V(\lambda^*, t) = f(\lambda^*, t)\left(e^{\lambda^* r} - 2\right) \sum_{t_j < t} \delta(t - t_j) \tag{7}$$

$$f(\lambda^*, t) = \frac{V(\lambda^*, t)}{\Omega(t)} \tag{8}$$

where the impulse $\delta(t-t_j)$ represents a spike at $t_j$ and $\Omega(t)$ is the normalizing denominator in Bayes rule (equation 2).

Equation 7 describes a first-order linear low-pass filter driven by impulses at event times of a Poisson process. Although it has been derived from principles of probability theory, it can be transformed into a biophysically realistic model of the subthreshold membrane potential of a neuron, simply by rewriting it using biophysically meaningful parameters of neuronal membranes (Shepherd, 1994; Koch and Segev, 1998),

$$C\frac{dv}{dt} + (v - v_L) g_L = (v - v_L) \omega \sum_{t_j < t} \delta(t - t_j) \tag{9}$$

where $v(t)$ is membrane voltage referenced to the leak current reversal potential or resting potential, $v_L$; $C$ is membrane capacitance; $g_L = C\lambda$ is leak conductance; and $\omega = \omega(\lambda^*) = C\left(e^{\lambda^* r} - 2\right)$ is the weight of the synaptic conductance $\omega\delta(t)$. By writing $v(t)$ on both sides of this equation we implicitly assume continuous renormalization, i.e. that some mechanism exists to implement the feedback loop specified by equation 8.

Equation 9 suggests that neurons could act as components of natural analog computers for Bayesian inference from spike trains.

## 3  Inference on a lattice

A population of neurons described by equation 9 with different membrane time constants, $\tau_k$, can compute the posterior density of $\lambda$ at a lattice of points $\lambda_k = 1/\tau_k$. The result of this computation is the membrane voltage of lattice neurons, interpreted as probability density at the corresponding lattice points. This requires some mechanism to continuously normalize membrane potentials across the lattice.

Normalization requires a fast feedback loop ensuring that total depolarization summed over lattice neurons is held constant (equation 8). This ensures that the membrane voltage of each individual neuron can be interpreted consistently as probability density at a point, even while currents due to incoming sensory spikes and leakage are fluctuating.

The required normalizing denominator can be approximated by

$$\Omega(t) = \sum_{k=1}^{N} v_k(t)\, d\lambda_k , \quad (10)$$

where $v_k(t)$ is the membrane potential of the $k$th lattice neuron and $d\lambda_k$ is a weight associated with the $k$th lattice point. In the simplest case, with uniformly spaced lattice points, this is an unbiased estimate of the integral in the denominator in equation 2 when the weights are equal and constant, $d\lambda_k = d\lambda$. This constant determines scaling of membrane potential to probability density, because $\Omega(t)$ is 1 when the lattice is normalized, that is when $v_k(t)d\lambda$ is a probability distribution for $\lambda_k$.

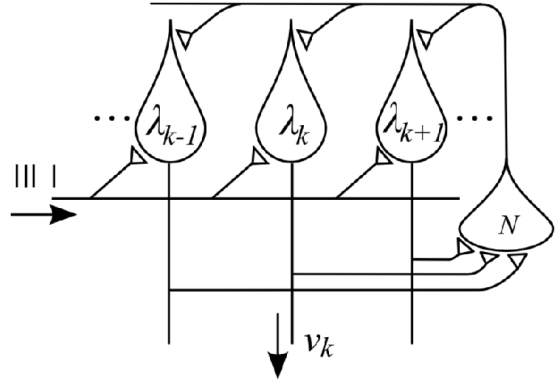

Figure 1: Anatomy of a Bayesian lattice filter for inferring the Poisson parameter of a spike train. Sensory spikes arriving from the left are distributed to lattice neurons. A pool of nonlinear neurons ($N$) form a feedback loop that continuously adjusts lattice neuron input gains. The membrane potential $v_k$ of a lattice neuron represents the posterior density of $\lambda$ at the lattice point.

Because membrane voltages are controlled via conductance changes, normalization requires controlling ion channel conductances such that the effect on membrane voltage is equivalent to dividing by $\Omega(t)$. A conductance

$$(\frac{1}{\epsilon} - 1) - \frac{(1/\epsilon)\Omega(t)}{\epsilon + \Omega(t)} \stackrel{\text{def}}{=} g_C + g_N \quad (11)$$

on each lattice neuron, where  is a small positive quantity, has the required effect (see Appendix).

The required conductance naturally decomposes into a constant term, $g_C$, and a dynamic term, $g_N$, that depends on total depolarization in the lattice. The constant term can be implemented by increasing the leak conductance of every lattice neuron ($g_L$ in equation 9) by $g_C$. The dynamic term can be implemented using a feedback loop around the lattice, containing a saturating nonlinearity of the form $g_N$. This adds a term on the right hand side of equation 9 that depends on total lattice activity.

Figure 1 shows a block diagram of a circuit that implements these equations. This circuit emerges as a natural solution to the problem of inferring the parameter of a Poisson process from an observed sequence of events, under the constraint that it must be done using a population of leaky integrators parameterized by input conductances. The solution is represented by membrane potentials of lattice neurons. The following section shows how the solution can be represented and transmitted using spikes.

# 4 Representation and inference with spikes

Spikes can be incorporated into the model by generating spikes in each neuron as an inhomogeneous Poisson process, with intensity proportional to its membrane voltage, $v(t)$.

The feedback loop can be implemented using a pool of spiking neurons, each having a short integration time constant comparable to the refractory period of the observed spike train, and a static membrane nonlinearity matching the second term in equation 11. By integrating convergent input from lattice neurons, each of these feedback neurons effectively counts spikes in the lattice over a small time window, and generates a spike in that time window with probability proportional to the magnitude of $g_N$. The expected number of spikes in a population of $N_N$ such neurons at any instant is $N_N g_N$. The required net conductance will be produced if feedback synapses onto lattice neurons each have conductance $1/N_N$. Because only the overall gain of the feedback loop matters, there is a family of equivalent solutions. For example, doubling the sensitivity of feedback neurons and halving their synaptic weights has no effect on average feedback conductance.

If lattice neurons generate Poisson spike trains with intensity proportional to their membrane voltage, then the feedback loop normalizes the expected number of spikes in the lattice during the integration time of the feedback neurons to $1/d\lambda$. The total probability mass of spikes in the normalized lattice then is 1. Each spike can be thought of as representing a possible state of the observed system, with a small probability $d\lambda$ attached to it.

The probability that there is a spike in any particular lattice neuron at any instant is proportional to the estimated posterior density of $\lambda$ at the corresponding lattice point. It follows that the set of lattice points corresponding to spiking lattice neurons is an approximate random sample from the posterior at every instant. Thus the network acts as a lattice particle filter, in which each spike is a particle. It generates a representation of the Bayesian posterior density of the Poisson parameter of an input spike train in real time, using neurons as operators and spikes as operands.

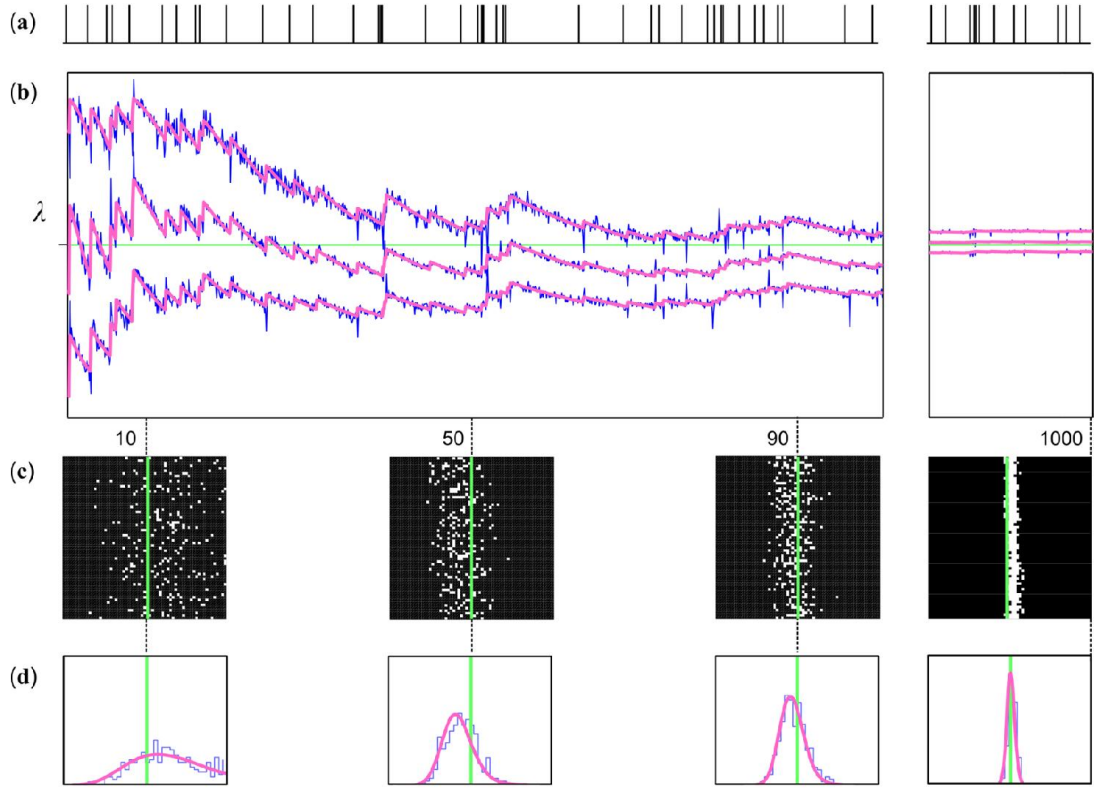

Figure 2: Simulation of a spiking neural lattice particle filter. (a) Observed spike train lasting 1000 seconds. First 100 and last 20 seconds are shown. (b) 10th, 50th and 90th percentiles of the posterior density, computed by numerical sequential Bayes (magenta), and by the neural lattice filter (blue). (c) Snapshots of spiking activity in the lattice at
10, 50, 90 and 1000 seconds. (d) Posterior density of $\lambda$ at snapshot times by sequential Bayes (magenta) with histograms of instantaneous lattice neuron spike counts (blue).

## 5   Example

Figure 2 shows plots generated by numerically solving equations 9 and 10, with additional feedback conductances in equation 9 as specified by equation 11. An explicit derivation of the discrete-time equations is provided in the Appendix. Communication between neurons in the model is mediated by spikes. Each neuron fires with probability proportional to its membrane voltage in a timestep equal to the refractory period, and current inputs to each neuron are computed by summing synaptic inputs to target neurons on the next timestep. The lattice in this example contains 4096 lattice neurons with lattice points equally spaced between 0 and 1. The feedback neurons have a nonlinearity with $\epsilon = 1/32$.

A sensory spike train is shown in panel (a). This is a sample from a censored Poisson process with a refractory period 5ms and constant intensity $\lambda$ = 0.5 spikes/s. The figure shows the first 100 seconds and the last 20 seconds of a spike train observed for 1000 seconds. In panel (b) the posterior has been constructed by a discrete-time spiking neural lattice particle filter. Percentiles of the posterior estimated by the lattice are shown in blue. The posterior has also been computed by solving Bayes rule (equation 2), sequentially in discrete time with a 5ms sample period. Percentiles of the true posterior computed this way are overlaid in magenta. This shows that while the posterior density converges quite slowly towards the true value $\lambda$ = 0.5, the spiking neural lattice particle filter estimates it precisely at all times.

Panel (c) shows spikes in 'snapshots' of the lattice, where the 4096 lattice neurons have been laid out in a 64x64 array. In panel (d) histograms of spike counts in each column of the array at snapshot times are superimposed on posteriors computed by numerical sequential inference. This confirms that the instantaneous distribution of spikes in the lattice accurately represents the true posterior in real time.

# 6   Translation: What in the world?

Determining the parameter of a Poisson process from a sample may not be a problem of interest to any animal with a nervous system, present company excepted. Decisions and actions in the real world require fast inferences about states and parameters of real systems. For example, a brain may observe a nearby animal via spikes from a sensory neuron firing with intensity $\lambda(\sigma(r))$ depending on the strength of the stimulus which in turn depends on distance to the source. A brain has no real interest in estimating its own sensory neuron's firing rate, $\lambda$, or in reconstructing the stimulus, $\sigma$, but it may well be interested in finding out what in the world is causing the sensory neuron to behave this way, $r$ (Figure 3).

The posterior density of $r$ can be obtained from the posterior density of $\lambda$ by a coordinate transformation. This transformation re-labels lattice points so that $\lambda_k$ maps to $r_k$ where $\lambda_k = \lambda(r_k)$ (Figure 3). The probability density at the lattice point is different in the new coordinates, but the volume of parameter space is stretched accordingly, and the net result is that the probability mass associated with a spike is unchanged (see Appendix). A spiking neural lattice particle filter constructed to infer $\lambda$ can simply be re-interpreted as a filter for inferring another parameter that influences $\lambda$.

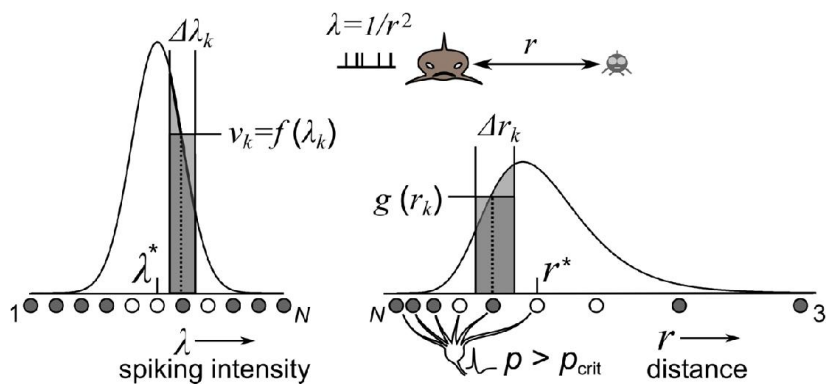

Figure 3: Real-world inference and decision-making. Distance $r = 1/\sqrt{\lambda}$ to a target can be inferred from the posterior density of $\lambda$ by re-labelling lattice points. Open circles represent spiking lattice neurons and filled circles represent inactive lattice neurons. The decision neuron sums probability mass in spikes at lattice points with $r_k < r^*$, or $\lambda_k > \lambda^*$, and fires if there is a sufficiently high probability ($p > p_{crit}$) that the target is closer than a critical distance $r^*$.

The equations for a spiking neural lattice particle filter do not require equally spaced lattice points or equally weighted spikes. It has been convenient to derive and illustrate the model with these assumptions because then it is possible to estimate probabilities simply by counting spikes. Also, when the geometry of the lattice matches the geometry of parameter space it is possible to visualize probability densities simply by visualizing the spatial density of spikes. Particle filters are generally more efficient with unequally weighted particles (Doucet et al., 2001). The practical significance of lattice spacing in the present context is a trade-off between acuity and the cost of neurons. In figure 3, acuity is uniform in Poisson parameter space, but higher for closer targets in the real world. This may be efficient because if an observer cares less about distant objects it should invest in fewer neurons to locate them. In general we might expect finer lattice spacing around critical parameter values.

# 7  Discussion

Historically, neuronal models like equation 9 were developed using empirical data from electrophysiology and molecular biophysics (Shepherd, 1994; Koch and Segev, 1998). Here the model has been derived by asking how a basic operation that underpins probabilistic inference can be implemented using mechanisms available to neurons. It turns out that a small, elegant network of spiking neurons can quickly, efficiently and accurately compute and represent Bayesian posterior densities and probabilities given sensory spike train data.

The spiking neural particle filter model described in this paper can infer the parameter of a Poisson process, or an underlying parameter such as distance to a signal source. This is perhaps simpler than any real inference problem faced by any extant animal. It may, however, have been crucial for survival of the earliest animals with nervous systems, in the late Ediacaran period. The emergence of carnivores at that time would have put a premium on information about proximity to other animals (Monk, 2014). The results reported here show that it is not difficult in principle for a small population of neurons to quickly and accurately infer distance to a predator or prey, given observations from a highly sensitive threshold-triggered proximity detector operating below the thermal noise floor.

If modern nervous systems employ the principles of representation and neural computation embodied in this model, then it is likely that evolution will have sculpted and elaborated them differently for different tasks. For example, it may pay to take a moment or two to decide whether an approaching object is a predator or prey, but dynamical state estimation for agile movement in relation to it must be done in real time. Different inference problems with different constraints, given different soures of sense data, may be solved in modern animals by neural networks as different from each other as cerebrum and cerebellum, even if they evolved from a common ancestral circuit, that perhaps performed a simple inference task that has no utility in a modern ecosystem.

The analysis reveals a remarkable correspondence between the mathematics of probabilistic inference and the physics of neuronal membranes. The normalized subthreshold membrane potential of a neuron (equation 9) is an exact analog computer for the numerator in Bayes rule for sequential inference given spikes generated by a homogeneous Poisson process. A specific kind of nonlinearity is predicted in feedback neurons. All neurons are predicted to have Poisson-like firing statistics. In a classical signal processing framework the sluggish, noisy, nonlinear response properties of neuronal membranes would appear to place constraints on the speed and accuracy of neural computation. But the Monte Carlo Bayesian framework employed here offers a new perspective: Neurons are not noisy, low-precision devices that perform accurate signal transformations by averaging over large populations, but high precision devices that perform distributed probabilistic computations exactly.

The model presented in this paper has been constructed using a criterion of of mathematical elegance consistent with known properties of neurons. It employs some abstractions and approximations whose biophysical implementation has not been spelled out in detail. For example we have simply assumed that neuronal spike trains can be modeled as inhomogenous Poisson processes, without modeling the biophysical implementation. Nor have we asked whether the particle filtering mechanism could be extended to deal with non-Poisson spike train data. Further detailed biophysical modeling will be required to examine if and how the proposed principles of representation and inference by spiking neurons could be implemented in real neural circuits.

The present model is not intended to model any particlar existing circuit, and whether its principles apply in any real nervous system remains to be examined. However, the proposed principles and mechanisms are consistent with neurobiology. Poissonlike firing statistics are common in spiking neurons (Richmond, 2009; Johnson, 2010). Nonlinearities of the required form are common (French, 2009; Silver, 2010; Priebe and Ferster, 2012), and synaptic gain control has been reported in electrophysiological and theoretical studies (Bastian, 1986; Nelson, 1994; Burkitt et al., 2003; Mehaffey et al., 2005; Rotem et al., 2007). Normalization of neuronal population activity is so widespread that it has recently been referred to as the prime example of a canonical neural computation (Carandini and Heeger, 2012). Thus all of the mechanisms required to implement a spiking neural lattice particle filter in the manner suggested here are known to occur in nervous systems, and other mechanisms are possible (Silver, 2010; Carandini and Heeger, 2012).

The 'labeled lattice' generalizes the 'labeled line' principle of neural representation, according to which the information content of a spike depends on where it originates in the nervous system (Shepherd, 1994). The spikes-as-particles concept shows that the common assumption that neural computation is naturally based on transformations of firing rates or firing rate vectors (Georgopoulos et al., 1986) is unnecessary and potentially misleading. Neurons can cut out the middle-man, and make inferences about the world and the body directly from spikes. Spike times and spike rates are merely times and rates at which signals are transmitted and representations are updated. Spikes are the operands of neural computation.

## Acknowledgments

Supported by a NZ Marsden Fund award and the NZ Performance-Based Research Fund. We thank Larry Hoffman, Chris Assad, Travis Monk and Helen Davies.

# Appendix

1. Feedback conductance for normalizing gain control

   Instantaneous normalization requires
   $$v(t+dt) = \frac{v(t)}{\Omega(t)} = v(t) + \frac{1-\Omega(t)}{\Omega(t)}v(t)dt.$$

   So the required conductance in continuous time is
   $$\frac{1-\Omega}{\Omega} = \frac{1}{\Omega} - 1 \approx \frac{1}{\epsilon+\Omega} - 1,$$

where $0 < \epsilon << 1$.

Now
$$\frac{1}{\epsilon + \Omega} - 1 = \frac{1}{\epsilon} - \left(\frac{1}{\epsilon} - \frac{1}{\epsilon + \Omega}\right) - 1$$
$$= \frac{1}{\epsilon} - \left(\frac{\epsilon + \Omega}{\epsilon(\epsilon + \Omega)} - \frac{\epsilon}{\epsilon(\epsilon + \Omega)}\right) - 1$$
$$= (\frac{1}{\epsilon} - 1) - \frac{(1/\epsilon)\Omega}{\epsilon + \Omega}$$

as required in equation 11. The analysis applies also in the discrete case, with $dt$ replaced by $\Delta t$. Inference is accurate and robust for values of between 0.01 and 0.1.

2. Construction of discrete time model for numerical simulations.

A refractory-censored Poisson process in continuous time is approximated by a Bernoulli process in discrete time, with the sample period of the discrete-time process equal to the refractory period of the continuous-time process. In discrete time an observed spike train is represented by a binary sequence

$$s_j = \begin{cases} 1 & \text{if a spike occurs} \\ 0 & \text{if no spike occurs in the} \end{cases}$$
$j$th sample period.

Let $\alpha_j$ be the probability that there is a spike in the $j$th sample period. To a good approximation,
$$\alpha_j = 1 - e^{-\lambda(j)r}$$

where $\lambda(j)$ is the average intensity of the Poisson process during the $j$th sample interval and $r$ is the sample period. Let $\lambda_k = k\lambda_{max}/N, k = 1,...,N$ be a regular lattice of points in the range of possible values of $\lambda$.

The likelihood of $\lambda_k$ given a spike in the $j$th interval is the probability of a spike given $\lambda = \lambda_k$ at that time,
$$L(\lambda_k|s_j = 1) = P(s_j = 1|\lambda = \lambda_k) = \alpha_j.$$

The likelihood of $\lambda_k$ given no spike in the $j$th interval is

Bayesian Inference with Spiking Neurons$$L(\lambda_k|s_j = 0) = P(s_j = 0|\lambda = \lambda_k) = 1 - \alpha_j.$$

These can be combined into a single expression for likelihood at the *j*th sample time given the sensory neuron state

$$L(\lambda_k|s_j) = (1 - \alpha_k) + (2\alpha_k - 1)s_j.$$

If the prior probability of $\lambda = \lambda_k$ at the $(j - 1)$th sample time is $v_k(j - 1)$, then the posterior probability at the *j*th sample time is given by Bayes rule,

$$v_k(j) = \frac{(1 - \alpha_k)v_k(j-1) + (2\alpha_k - 1)v_k(j-1)s_j}{\sum_{k=1}^{N} L(\lambda_k|s_j)v_k(j-1)d\lambda}$$

This equation computes the posterior density of $\lambda$ on a lattice sequentially at sample times, given a spike train. It is implemented in the MATLAB model as a two-stage cascade at each time step. In the first stage the numerator is evaluated for every *k*, and spikes are generated to form a non-normalized sample. In the second stage the weighted sum of spikes is fed back to dynamically adjust conductances on lattice neurons at the next time step. By operating within a sample interval, the feedback loop is fast enough to stably normalize the lattice.

3. Particle mass is not affected by change of variables

Suppose $\lambda$ has probability density $f(\lambda)$ and $\lambda(r)$ is a smooth function of *r*. Then each lattice point $\lambda_k$ transforms to $r_k$ such that $\lambda(r_k) = \lambda_k$. In the transformed density function, $g(r)$, the probability mass associated with a particle at the lattice point is unchanged in the limit of high particle density

$$f(\lambda_k|S)d\lambda_k = g(r_k|S)dr_k.$$

and therefore

$$f(\lambda_k|S)\Delta\lambda_k \approx g(r_k|S)\Delta r_k$$

provided that $\Delta\lambda_k$ and $\Delta r_k$ are small. This is illustrated in figure 3. As discussed in the main text, "small" is determined by the ecological utility of sensory acuity. The analysis simply says that a brain can attain whatever acuity is worth paying for.